\documentclass[prb,aps,showpas,floatfix,twocolumn]{revtex4}
\usepackage{epsfig} \usepackage{graphics} \usepackage{bm} \usepackage{amssymb}
\begin{document}
\title{Quantum pumping of electrons by a moving modulated potential}
\author{Markku~J{\"a}{\"a}skel{\"a}inen}
\email{mrq@phy.stevens.edu}
\author{Frank Corvino}
\author{Christopher~P.~Search}
\affiliation{Department of Physics and Engineering Physics,
Stevens Institute of Technology, Castle point on the Hudson, Hoboken, NJ 07030, USA}
\author{Vassilios Fessatidis}
\affiliation{Department of Physics, Fordham University, Bronx, NY 10458, USA}
\date{\today}
\begin{abstract}
Quantum pumping holds great potential for future applications in micro- and nanotechnology. Its main feature, dissipationless charge transport, is theoretically possible via several different mechanisms. However, since no unambiguous verification has been demonstrated experimentally, the question of finding a viable mechanism for pumping remains open. Here we study quantum pumping in an one dimensional electron waveguide with a single time-dependent barrier. The quantum pumping of electrons using a potential barrier whose height and position are harmonically varied is analyzed analytically and by numerically solving the time-dependent Schr{\"o}dinger equation. The pumped charge is modeled analytically by including two contributions in linear response theory. First, the scattering of electrons off a potential moving slowly through matter-waves gives a contribution independent of the translational velocity of the potential. Second, Doppler-shifted scattering events give rise to a velocity dependent contribution, which is found in general to be small in comparison with the first one. The relative phase between the oscillations of the height and position is found to be the factor that determines to what extent either contribution is present.
\end{abstract}
%
%
%
%
%
%

\maketitle

\section{Introduction}\label{Introduction}
Quantum pumping \cite{Altshuler} is a novel way of transporting charge or spin\cite{Kunal} without applying bias voltages in nanoscale conductors. The main idea has been around for some time, beginning with the seminal work by Thouless \cite{Thouless}, who envisioned transport of charge in a moving periodic potential with similar results later obtained by Niu \cite{Niu}. The essential idea of pumping is that the electrons interact with a potential that depends on at least two independent parameters that vary periodically in time. When these parameters vary out of phase with each other, a finite dc current is produced that depends only on how the parameters are varied. If the cyclic variation of the parameters is much slower than all other time scales, the wave function of the electrons is adiabatically deformed and because of this quantum pumping is often called adiabatic quantum pumping. True quantum pumping is qualitatively different from classical dissipative rectification of an ac signal with the nearest classical analogue to quantum pumping being a peristaltic pump or Archimedean screw.  The first claim for experimental observation of quantum pumping was reported by Switkes \textit{et.al.}\cite{Switkes} where the shape of an electrostatically defined quantum dot was cyclically deformed. Theoretical work showed that time dependence of the experimental parameters may introduce stray capacitances that produce a rectifying effect \cite{BrouwerRect}, which may overshadow the contribution from quantum pumping. The validity of this scenario was later verified experimentally \cite{Marcus}.

A major objective in theoretical studies is to calculate the amount of charge transported per driving cycle for periodic signals. Thouless showed in his original work that the transported charge is quantized if the Fermi energy lies in an energy gap of the Hamiltonian \cite{Thouless} and others have shown that quantization of the pumped charge occurs due to Coulomb blockade \cite{levninson, andreev} although quantum pumping is not necessarily quantized \cite{Brouwer}.
B{\"u}ttiker, Thomas, and Pretre \cite{BTP} derived an expression  for current partition in multi-probe conductors, and expressed the charge due to quantum pumping in terms of the instantaneous scattering matrix and its derivatives with respect to the driving parameters. Brouwer\cite{Brouwer} used these results to derive a connection with geometric transport, where the adiabatic curvature measures the sensitivity of the quantum states to parametric changes in the Hamiltonian.
These results are all based on linear response theory. B{\"u}ttiker and Moskalets \cite{Moskalets} and also Kim \cite{FloquetKim} applied the technique of Floquet scattering to deal with situations beyond the linear response regime for periodic variations.

Most theoretical discussions of quantum pumping focus on either shape deformations or modulated tunneling rates of a quantum dot \cite{Brouwer,levninson,andreev,mucciolo} or variations of the amplitude of two localized potential barriers in a quantum wire \cite{Moskalets, FloquetKim}. The contribution to the pumped charge from a scatterer translated a finite distance was first discussed by Avron, Elgart, Grag, and Sadun \cite{AvronSnowPlow}. Cohen, Kottos, and Schanz \cite{CohenSnowPlow} treated translation in annular geometries, where magnetic fluxes give rise to Aharonov-Bohm type effects. However, periodic variations of the position alone of a scatterer in an open one dimensional waveguide will of course not produce any net pumped charge. In this paper we consider the quantum pumping by a barrier undergoing periodic translation together with the simultaneous modulation of its height. We thus extend earlier studies to the case of pumping with a single, localized barrier. We analyze the system by using both the formalism developed by B{\"u}ttiker, Thomas, and Pretre \cite{BTP}, and Brouwer\cite{Brouwer} and also by extending the results of Avron, Elgart, Grag and Sadun for translated potentials. The parametrically varied scattering matrix is taken as a starting point to derive results for both contributions. We argue that there are two mechanisms that contribute to the total pumped charge. The first is the "snow plow" dynamics of Avron, Elgart, Grag and Sadun resulting from pushing the electrons. The second is the Doppler shifted scattering of the matter waves off the potential that originates from the finite velocity of the potential.

Time-dependent studies of the quantum behavior of electrons in guided  nanostructures is relatively new. Fy and Willander\cite{Willander_WP} considered the effects of gate bias and device geometry on the I-V characteristics beyond a plane-wave model. Of more relevance to the results presented here is the work by Agarwal and Sen\cite{Agarwal}, where quantum pumping was studied in the time-domain for a tight-binding model. Oriols, Alarcon, and Fernandez-Diaz\cite{Oriols} studied the dynamics of independent electrons in phase-coherent devices beyond periodic driving with quantum pumping as an example. Therefore in addition to our analytic results, we study the quantum dynamics of the proposed pumping device by performing numerical simulations. These simulations allow us to visualize the effect of the pumping potential on the electron wave function starting from an empty wire.

The paper is organized as follows: In section \ref{Model}, we present our model and derive expressions for the two contributions to the pumped charge. In section \ref{Simulations}, we present numerical results and analyze the different contributions. Finally, in section \ref{Summary}, we summarize our results and discuss the implications of them.

\section{Theory}\label{Model}
Our aim here is to investigate the quantum pumping of a single translated and modulated potential barrier in a one dimensional quantum wire, both by using a plane wave scattering approach, and later in section \ref{Simulations} using numerical simulations. For simplicity we choose a specific model system basic enough to treat analytically, yet general enough to draw universal conclusions from. To achieve this, we study the quantum dynamics of electrons scattering off of the potential
\begin{equation}\label{Potential}
V(x,t)=\frac{V_0(t)}{\cosh^2[(x-x_c(t))/L]},
\end{equation}
where $L$ determines the width of the barrier. Both the position $x_c$ and the barrier height $V_0$ are varied harmonically around their central values with a difference $\Delta$ in relative phase. The amplitude is taken to oscillate around a central value, which changes according to
\begin{equation}\label{Amplitude}
V_0(t) = A_0[1+\kappa\sin(\omega t+\Delta)],
\end{equation}
and the barrier center is taken to depend on time as
\begin{equation}\label{PosAmplitude}
x_c(t) = x_0\sin(\omega t).
\end{equation}
Equation (\ref{PosAmplitude}) gives us for the instantaneous velocity of the barrier
\begin{equation}\label{PotVelocity}
v_c(t) \equiv \dot{x}_c(t) = \omega x_0 \cos(\omega t).
\end{equation}
Our choice of potential is due to convenience as Eq.~(\ref{Potential}) has an analytically known solution for the scattering matrix\cite{Guechi}.
\begin{figure}[ht]
\begin{center}
\epsfig{file=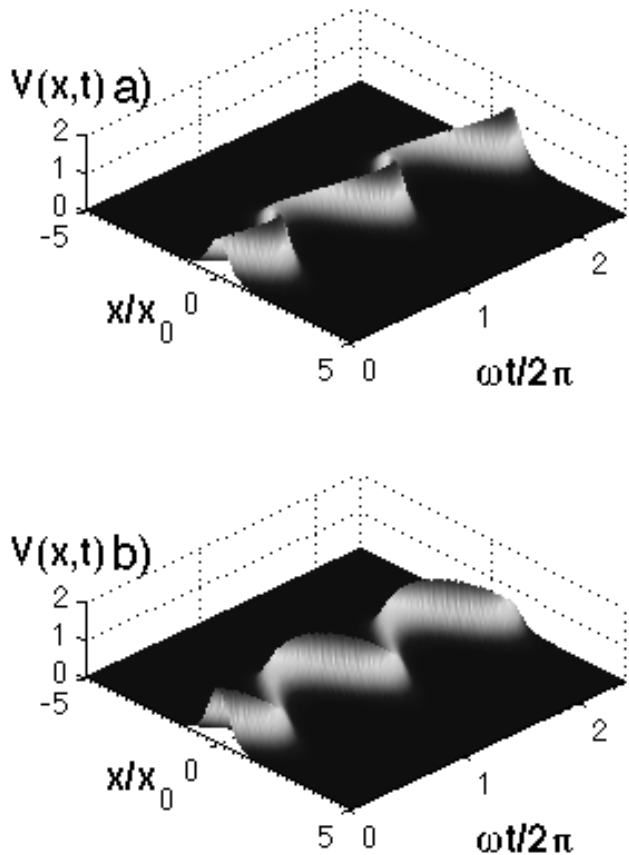,width=8.5cm}\caption{\label{PotSurfaces}The potential barrier (\ref{Potential}) as a function of time for (a) $\Delta = 0$, and (b) $\Delta = \pi/2$. Here $x$ is measured in units of $x_0$, which is defined in the text, $t$ is in units of $T$, $V$ is in units of $A_0$, and $\kappa=0.5$ while $L=x_0/2$.
For the case $\Delta = \pi/2$, which is shown in b), the potential always takes on values at or below $A_0$ as is it moves towards larger values of $x$, and similarly values larger than $A_0$ for the other half of the cycle. From this it follows intuitively that the 'snow-plow' mechanism can pump charge through quantum scattering of electrons.}
\end{center}
\end{figure}
In Fig.~\ref{PotSurfaces} the potential (\ref{Potential}) is shown as a function of time and space. In (a) the relative phase is $\Delta = 0$, and the potential reaches its highest and lowest values at the turning points of the trajectory. For this case, the barrier height changes most rapidly at $x = 0$. In (b) we have $\Delta = \pi/2$ and the extreme values of the height occur at $x = 0$, which is where the translational velocity peaks.
For the potential (\ref{Potential}), the scattering amplitudes for plane waves of momentum $k_F$ are given by \cite{Guechi}
\begin{equation}\label{r_coeff}
r(k_F,V_0)=\frac{\Gamma(1+\nu-ik_FL)\Gamma(-\nu-ik_FL)\Gamma(ik_FL)}{\Gamma(1+\nu)\Gamma(-\nu)\Gamma(-ik_FL)},
\end{equation}
and
\begin{equation}\label{t_coeff}
t(k_F,V_0)=\frac{\Gamma(1+\nu-ik_FL)\Gamma(-\nu-ik_FL)}{\Gamma(1-ik_FL)\Gamma(-ik_FL)},
\end{equation}
and where
\begin{equation}
\nu=\frac{1}{2}\left[-1+\sqrt{1-8V_0L^2}\right].
\end{equation}
$\Gamma(z)$ is the standard Gamma function. As the potential (\ref{Potential}) is translated along the x-axis in an oscillatory manner, as shown in Fig.~\ref{PotSurfaces}, particles which are scattered off of it see transmission/reflection amplitudes modulated periodically in time. We note here that the scattering matrix does not depend explicitly on the position $x_c(t)$ of the scatterer. Rather the change of position gives rise to a pumped net charge in two physically distinct ways.

\begin{figure}[ht]
\begin{center}
\epsfig{file=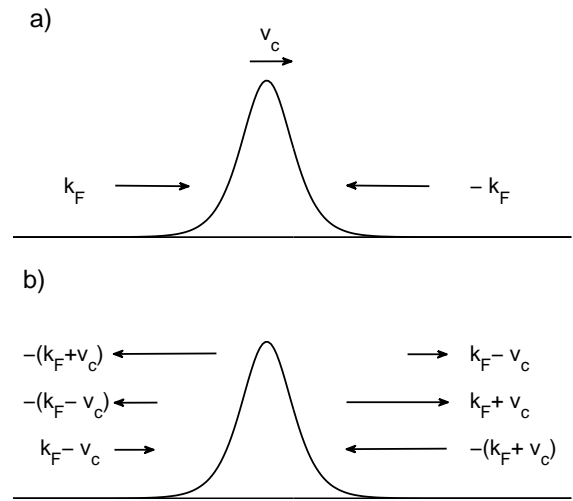,width=8.5cm}\caption{\label{Sketch}The scattering of electrons off the potential (\ref{Potential}) shown schematically in a) the lab frame and b) an inertial frame at rest relative to the potential. The momenta of the incoming and scattered waves as well as the instantaneous velocity of the potential are shown by arrows whose lengths are proportional to the magnitude of the velocities. In the lab frame shown in (a) the incoming plane waves have momenta of equal magnitude given by $k_F$.
In the moving frame shown in (b) the left going waves have larger momenta than the right going ones, as indicated in the figure. In the moving frame the potential is stationary and the scattering can be treated using standard scattering theory with the modification that the momenta of the left and right going plane waves are shifted as indicated above. In (b) the momenta correspond from top to bottom to: transmitted, reflected and incoming waves respectively.}
\end{center}
\end{figure}

First, a contribution to quantum pumping occurs for any finite period due to Doppler shifting of the reflection and transmission amplitudes of the potential. This can be understood intuitively by considering the scattering off of the moving barrier from an inertial frame at rest with the scattering potential. In a frame moving at velocity $v_c$, the instantaneous velocity of the potential, we find that the potential appears to be stationary, but that the momenta of plane waves propagating at $k=\pm k_F$ change to $k=\pm k_F-v_c$. A pedagogical sketch of this is shown in Fig.~\ref{Sketch}, where the potential is shown together with vectors representing both the velocity of the potential and the momenta of plane waves propagating inwards towards the potential. In (a) the velocities are shown in the lab frame, and the momenta of the incoming plane waves are equal in magnitude. In (b) the same situation is shown in a frame moving with $v_c$ to the right now also including the scattered waves. All momenta are now shifted by the translational velocity of the potential, and an asymmetry between positive and negative momenta is created.
In the moving frame the potential is stationary and the scattering can be treated using standard scattering theory with the modification that the momenta of the left and right going plane waves are shifted as indicated in Fig.~\ref{Sketch}.


Applying this to the scattering of plane waves, we find from Galileo invariance of the Schr{\"o}dinger equation that the instantaneous scattering matrix, which relates the incoming and outgoing amplitudes, is given by a $4\times 4$ S-matrix corresponding to the propagating modes $\pm k_F-v_c$ and $\pm k_F+v_c$.
The reflection and transmission probability amplitudes in the moving frame are $\tilde{r}_{\pm}=r(k_F\pm v_c,V_0)$ and $\tilde{t}_{\pm}=t(k_F\pm v_c,V_0)$. Returning to the laboratory frame, the transmitted waves have momentum $\pm k_F$ while the reflected waves have momentum $\pm k_F +2v_c$, thus being scattered inelastically, and we find that the corresponding non-zero scattering probabilities in the S-matrix are given by (for $k_F>2v_c$) \cite{Moskalets,FloquetKim,Atabek,datta}
\begin{eqnarray}
|S_{-k_F\rightarrow -k_F}|^2&=&|t_+|^2=\frac{k_F}{k_F}|\tilde{t}_{+}|^2, \\
|S_{k_F\rightarrow k_F}|^2&=&|t_-|^2=\frac{k_F}{k_F}|\tilde{t}_{-}|^2, \\
|S_{k_F\rightarrow -k_F+2v_c}|^2&=&|r_-|^2=\frac{k_F-2v_c}{k_F}|\tilde{r}_{-}|^2, \\
|S_{k_F\rightarrow k_F+2v_c}|^2&=&|r_+|^2=\frac{k_F+2v_c}{k_F}|\tilde{r}_{+}|^2.
\end{eqnarray}
In this case we thus explicitly take into account the velocity dependence of the scattering against a moving target. This results in different scattering energies for the electrons when considered from a frame at rest with the barrier, and can be viewed as pumping due to Doppler-shifted scattering events.
We note here that as a consequence of the translational motion, Eqs.~(\ref{r_coeff},\ref{t_coeff}) are valid only when
\begin{equation}\label{TunnCond}
\frac{(k_F+\omega x_0)^2}{2}< A_0(1-\kappa),
\end{equation}
where $k_F+\omega x_0$ is the maximal instantaneous momentum of the plane wave in the moving frame. Equation (\ref{TunnCond}) is a condition for the dynamics to take place in the tunneling regime, when the total kinetic energy in the moving frame is smaller than the potential height at its minimum.

For any two-terminal device where two independent parameters $X_1$ and $X_2$ are varied cyclically, the charge accumulated per period on the left/right lead  is given by\cite{Brouwer}
\begin{equation}\label{Q_Brouwer}
Q_{L,R}=\frac{e}{\pi}\int_S\sum_{i,j\in L,R} Im\frac{\partial S^{*}_{ij}}{\partial X_1}\frac{\partial S_{ij}}{\partial X_2}dX_{1}dX_{2},
\end{equation}
where the label j stands for the modes propagating towards the left (L) or right (R) lead respectively.
Here, using the elements of the scattering matrix and Eq.~(\ref{Q_Brouwer}), we obtain for the charge per cycle on the right lead due to the Doppler shifted contribution
\begin{widetext}
\begin{equation}\label{Q_Surf}
Q_D=\frac{e}{\pi}Im\int_{S}(
\frac{\partial t^{*}_{-}}{\partial v_c} \frac{\partial t_{-}}{\partial V_0}-
\frac{\partial t_{-}}{\partial v_c} \frac{\partial t^{*}_{-}}{\partial V_0}+
\frac{\partial r^{*}_{+}}{\partial v_c} \frac{\partial r_{+}}{\partial V_0}-
\frac{\partial r_{+}}{\partial v_c} \frac{\partial r^{*}_{+}}{\partial V_0}
)dS.
\end{equation}
\end{widetext}
Alternatively, the charge can be calculated from
\begin{equation}\label{Q_Line}
Q_D=\frac{e}{\pi}Im\int_{\partial S}(t^*_-\nabla t_-+r^*_+\nabla r_+)\cdot d\vec{X},
\end{equation}
where we have introduced the notation $\vec{X}=(v_c,V_0)$, and $\nabla=(\partial/\partial v_c,\partial /\partial V_0)$.
For the case of strong pumping, i.e. when the integrand in Eq.~(\ref{Q_Surf}) varies appreciably over the integration area, the contribution is easier to calculate numerically using Eq.~(\ref{Q_Line}). The reason for this simply being that a one-dimensional integral requires less computation time than a two-dimensional one, and the increase in computation time for using Eq.~(\ref{Q_Surf}) becomes noticeable for rapidly varying integrands. Both expressions Eq.~(\ref{Q_Surf}) and Eq.~(\ref{Q_Line}) depend on the derivatives of the reflection and transmission amplitudes, with respect to both the barrier height and to the translational velocity of the potential, and are complicated enough to prohibit the derivation of compact analytical expressions for the pumped charge.

A second contribution comes from moving a scatterer slowly an infinitesimal distance $dx_c$ through an impinging matter-wave of momentum $k_F$, and gives rise to pumping through a quantum mechanical ``snow-plow dynamics'' \cite{AvronSnowPlow}
\begin{equation}\label{dQ_SP}
dQ_{SP}=-\frac{e k_F}{\pi}|r(k_F,V_0)|^2dx_c,
\end{equation}
that can be interpreted as resulting from reflecting a fraction $|r|^2$ of the $k_Fdx_c/\pi$ electrons occupying the region in front of the barrier.
As discussed in Ref. [15], Eq.~(\ref{dQ_SP}) is obtained from the results of B{\"u}ttiker, Thomas, and Pretre\cite{BTP} for the adiabatically pumped charge due to moving the scatterer a distance $dx_c$.
The net transferred charge over one pumping cycle starting at an arbitrary time $t_0$ is then given by
\begin{equation}\label{Q_SP}
Q_{SP}(t_0)= -\frac{e k_F}{\pi}\int_{t_0}^{t_0+T}|r(k_F,V_0(t))|^2v_c(t)dt.
\end{equation}
The integrand of Eq. (\ref{Q_SP}) is explicitly dependent on time and on the velocity $v_c(t)$. Despite this, the pumped charge is independent of the velocity and its temporal evolution since $Q_{SP}$ is an adiabatic invariant. The occurrence of the velocity and the time only serve to parameterize the integration.
For the case $\Delta = 0$, illustrated in Fig.~\ref{Potential} (a), we expect that $Q_{SP} = 0$, since contributions coming from the potential moving to the right, when $v_c > 0$ are exactly canceled when the potential moves back to the left with $v_c < 0$ and identical value of the barrier height.
For $\Delta>0$, when the potential is moving to the left, the amplitude is on average below its central value, $A_0$, whereas during the motion in the other direction it is above the central value $A_0$ on average. The reflection probability is therefore larger when the potential moves to the right versus the left. This implies that a net charged will get pushed towards the right during one complete cycle. The difference in the amplitudes is maximal for  $\Delta = \pi/2$, and as a result the difference in charge reflected to the left and to the right is also maximal.

\begin{figure}[ht]
\begin{center}
\epsfig{file=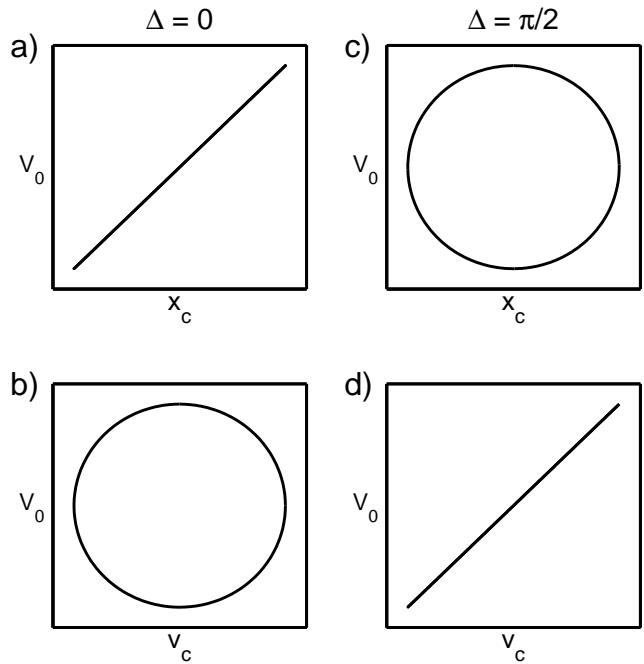,width=8.5cm}
\caption{\label{Lissajous} Integration contours in the parameters
$(x_c,V_0)$ and $(v_c,V_0)$ for the different phase values $\Delta = 0$ and $\Delta = \pi/2$. In (a) and (b) we have $\Delta = 0$, and we see that  the area enclosed in $(x_c,V_0)$ equals zero, whereas the area in $(v_c,V_0)$ is maximal. For $\Delta=0$ we thus expect that $Q_{SP}=0$, whereas $Q_{D}$ should be maximal. In (c) and (d) we have $\Delta = \pi/2$ and the situation is the opposite.}
\end{center}
\end{figure}

When $\Delta = \pi/2$, as shown in Fig.~\ref{Potential} (b), the rate of change of the height is  maximal for $x=\pm x_0$, where the translational velocity is equal to zero. Likewise, we have maximal velocity at $x=0$ when the change in height equals zero. For $\Delta=0$ this is exactly reversed.
The behavior of the relevant parameters is illustrated schematically in Fig.~\ref{Lissajous}, where the integration contours in $(x_c,V_0)$ and $(v_c,V_0)$ are shown for the different phases $\Delta = 0$ and $\Delta = \pi/2$. In (a) and (b) we have $\Delta = 0$, and we see that  the area enclosed in $(x_c,V_0)$ equals zero, whereas the area in $(v_c,V_0)$ is maximal. For $\Delta=0$ we thus expect that $Q_{SP} = 0$, whereas $Q_D$ is maximal. In (c) and (d) we have $\Delta = \pi/2$ and the situation is reversed so that we expect $Q_{SP}$ to be maximal while $Q_D=0$. Since of the two values $\Delta = 0$ maximizes the velocity dependent contribution and $\Delta = \pi/2$ maximizes the position dependent pumped charge, we can view the pumping in this system as being the combination of two physically different mechanisms whose relative contributions are determined by the choice of relative phase between the driving parameters. The pumped charge for $\Delta=-\pi/2$ and $\pi$ will be the same as for $\Delta=\pi/2$ and $0$, respectively, except that the direction of the pumped charge is reversed.
We also note here that the contribution from the Doppler shifted scattering depends on the velocity of the scatterer, and thus goes to zero in the limit of infinitely slow driving for finite spatial amplitude $x_0$, leaving only $Q_{SP}$ given by Eq.~(\ref{Q_SP}) as a contribution to the pumped charge.

\section{Simulations}\label{Simulations}

\begin{figure}[ht]
\begin{center}
\epsfig{file=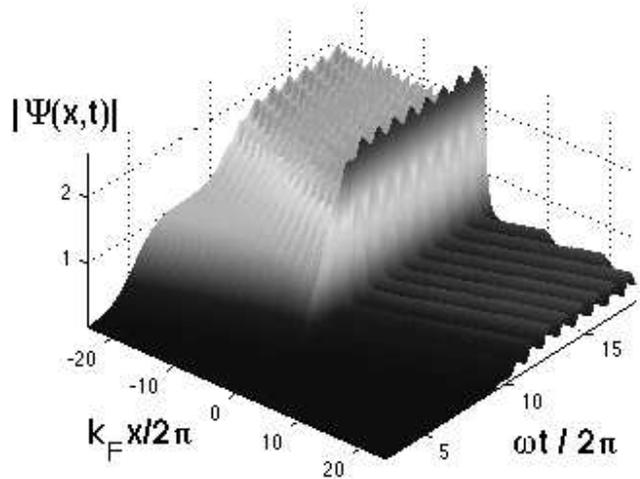,width=8.5cm}\caption{\label{Transient} The quantum dynamics of an initially empty scattering region. Matter-waves are seen to enter the region from a source on the left, and after an initial transient a periodic stable state is reached. The oscillating peak at $x=0$ is due to the interference in the scattering region where the parts to be transmitted still overlaps with the incoming and reflected parts. Note that $x$ is in units of the Fermi wavelength, and $t$ is in units of $T$. For the potential we used the following values $k_FL/2\pi=3/\pi$, $2A_0/k_F^2=10/9$, $\kappa=0.05$, and $k_F x_0/2\pi = 3/40\pi$.}
\end{center}
\end{figure}

To investigate the presence of quantum pumping in our system in more detail, we simulate the dynamics by solving the following time-dependent Schr{\"o}dinger equation numerically\cite{WavePackets}
\begin{equation}\label{TDSE}
i\frac{\partial\Psi}{\partial t} = -\frac{1}{2}\frac{\partial^2\Psi}{\partial x^2} +V(x,t)\Psi + S(x,t),
\end{equation}
where $V(x,t)$ is given by Eq.~(\ref{Potential}), and $S(x,t)$ a function describing the source, here taken to be both phase coherent and quasi-monochromatic,
\begin{equation}\label{Source}
S(x,t) = iS_0\exp(-(x-x_s)^2/L_s^2\pm i k_F x),
\end{equation}
where $S_0$ is the source strength, $x_s$ is the central position of the source, and $L_s$ is the width of the source.
For numerical convenience, complex absorbing potentials \cite{CAP} were used to implement transparent boundaries and thus avoid any significant effects of finiteness of the numerical grid on the dynamics. Note that here, as in the rest of the paper, we work in units where $\hbar=m=1$. Figure \ref{Transient} shows the dynamics of a typical simulation where charge is injected from the left lead into an initially empty scattering region.
An initial transient when the scattering region is filled by the matter-waves is seen to be followed by a periodic regime. To the right, for $x > 0$, trains of transmitted waves are seen to exit the scattering region  with amplitudes modulated in time. To the left, for $x < 0$, the reflected part is seen superimposed on, and interfering with the incoming waves.

For a more quantitative check, we calculate the pumped charge in the time-domain  by integrating the difference in the instantaneous probability currents over one period of time. These are measured at two points $x_{\pm}$, chosen  at distances sufficiently far away from the scattering region that represent the outgoing leads, and we have
\begin{equation}\label{Q_WP}
Q(t) = e\int_t^{t+T}\left[J_+(x_+,t')-J_-(x_-,t')\right]dt',
\end{equation}
where $J_{\pm}$ is the total probability current of electrons being in the lead to the right (left) of the barrier.
In any realistic implementation of quantum pumping of electrons, the two leads are connected to independent reservoirs with effectively no phase coherence between electrons injected from either one. To account for this, the two currents $J_{\pm}$ were calculated by having the sources placed in opposite leads, and the charge imbalance was calculated as the incoherent difference by using completely independent numerical simulations.
We also note here that the agreement between the simulations of quantum dynamics of scattering with results calculated using the corresponding instantaneous values of the scattering parameters for plane waves only will agree if the momentum distributions used are narrow enough, as discussed by Atabek and Lefebvre\cite{Atabek}.
\begin{figure}[ht]
\begin{center}
\epsfig{file=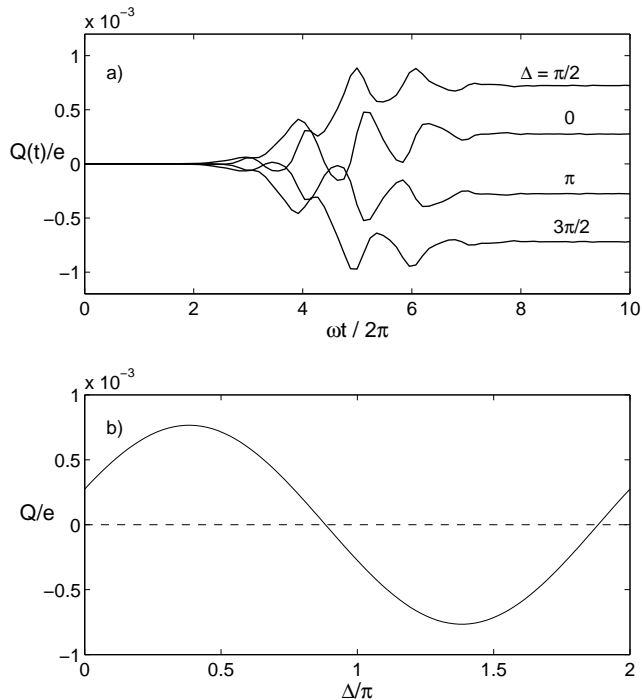, width=8.5cm} \caption{\label{Delta}In (a) The pumped charge (in units of $e$) as a function of time (in units of $T$) is shown for four different values of the relative phase $\Delta$. In (b) the  value of $Q$ is shown as a function of $\Delta$ for large values of $t$. We note here that the shape of the curve in (b) indicates that the pumped charge is basically adiabatic in the sense that the dependence as a function of $\Delta$ is harmonic, and the pumping exhibits an area dependence expected from a sum of parametric contributions as indicated in Fig.~\ref{Lissajous}. On the other hand, the presence of a nonzero contribution for $\Delta = 0$ indicates that the truly adiabatic limit where the pumped charge is independent of the velocity of the parametric changes, is not yet reached. For the potential we used the following values $k_FL/2\pi=3/\pi$, $2A_0/k_F^2=10/9$, $\kappa=0.05$, and $k_F x_0/2\pi = 3/40\pi$. Note that the pumping period was taken as $T=2\pi$ in units of $A_0^{-1}$ in these simulations.}
\end{center}
\end{figure}

In Fig.~\ref{Delta}, the resulting pumped charge is shown as a function of time and relative phase. The charges were averaged over one period of time and calculated from a set of simulations with different values for $\Delta$. In a) the net charge as a function of time is shown for four different values of the relative phase. For all values there is a transient behavior followed by an asymptotic stationary value. In b) the pumped charge in the asymptotic regime is shown as a function of the relative phase $\Delta$. The fact that the charge is nonzero for both $\Delta = 0$ and for $\Delta = \pi/2$ shows that there are two distinct contributions to the dynamics, each dependent on the parameter combinations ($x_c(t),V_0(t)$) and ($v_c(t),V_0(t)$) respectively, as discussed in section \ref{Model}.

\begin{figure}[ht]
\begin{center}
\epsfig{file=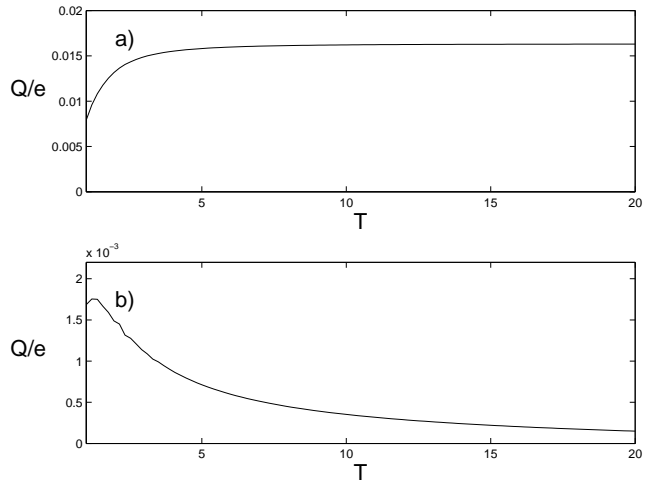,width=8.5cm}\caption{\label{Adiab}Pumped charge (in units of $e$) as a function of the oscillation period $T$ (in units of $1/A_0$). In (a) $\Delta = \pi/2$ we see that for large values of the period the charge approaches a constant value larger than zero, as expected from Eq.~(\ref{Q_SP}). In (b) $\Delta=0$ the pumped charge approaches zero for large periods, in accordance with Eq.~(\ref{Q_Line}) since $max(v_c) = \omega x_0\propto 1/T$. For the potential we used the following values $k_FL/2\pi=3/\pi$, $2A_0/k_F^2=10/9$, $\kappa=0.05$,  and $k_F x_0/2\pi = 3/40\pi$.}
\end{center}
\end{figure}

We expect the system to behave adiabatically for slow enough driving\cite{Elgar}, which is where the instantaneous scattering matrix used in Sec. II should describe the pumping well. For the snow plow contribution, which only depends on the position and amplitude, not the velocity, we expect from adiabaticity that the pumped charge over one cycle will reach a steady value for large pumping periods. In Fig.~\ref{Adiab} the pumped charge is shown as a function of the pumping period $T$ for the phases $\Delta = \pi/2$ in (a), and $\Delta = 0$ in (b). In Fig.~\ref{Adiab} (a) we see that for large values of $T$, a constant value is reached, as we expect from our analysis. For $\Delta = 0$, on the other hand, we have that the area of the integration domain in Eq.~\ref{Q_Surf} scales with $max(v_c) = \omega x_0 \propto 1/T$, and we thus expect that the pumped charge approaches zero, as is also seen in Fig.~\ref{Adiab} (b), where $Q(T)\propto 1/T$ for large values of $T$.
For small (non-adiabatic) periods, the pumped charge obtained from the simulations deviates from what we expect using Eqs.~(\ref{Q_Line}) and (\ref{Q_SP}), and for both cases the general result is that we find smaller values of the total pumped charge. In Fig.~\ref{Adiab} (a) this is seen as the value in the simulation drops below the constant asymptote, and in (b) the values from the simulation fall below the $1/T$ behavior expected from the scaling of Eq.~(\ref{Q_Line}).

\begin{figure}[ht]
\begin{center}
\epsfig{file=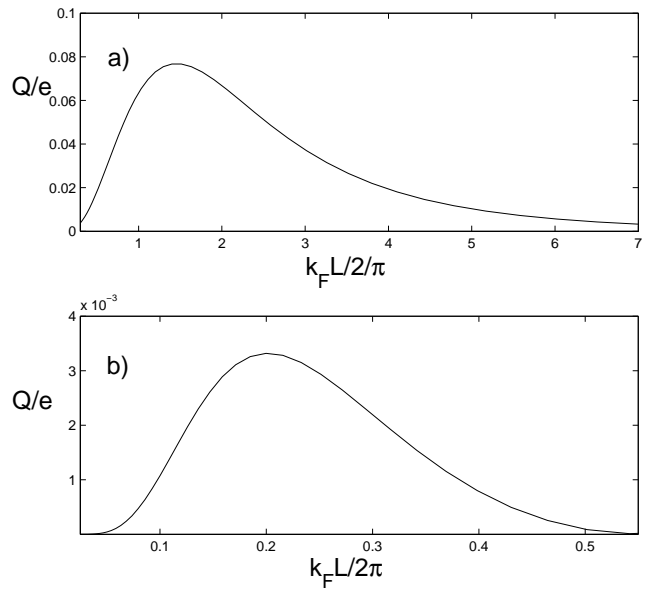,width=8.5cm}\caption{\label{Fig:Tunn}Pumped charge (in units of $e$) as a function of barrier width for two different values of the relative phase. In (a) we have $\Delta = \pi/2$, and in (b) $\Delta = 0$. For both cases the pumped charge is seen to have a maximum as a function of barrier width.  For the scattering potential the following values were used $T=2\pi A_0^{-1}$, $2A_0/k_F^2=10/9$, $\kappa=0.05$, and $k_F x_0/2\pi = 3/40\pi$.}
\end{center}
\end{figure}

For small values of L, the barrier becomes increasingly transparent, and as a result the integrand in Eq.~(\ref{Q_SP}) decreases. On the other hand, for very large widths, the reflection coefficient becomes nearly unity, nearly independent of the translational velocity or barrier height, so that the charge pushed to the right during one half-cycle exactly cancels the charge pushed to the left during the second half-cycle. From these two limits we deduce that there is a maximum in the pumped charge somewhere at moderate values of the width. In Fig.~\ref{Fig:Tunn} the pumped charge is shown as function of $L$, the barrier width, for two different values of the relative phase, in (a) we have $\Delta = \pi/2$, and in (b) $\Delta = 0$. For both cases the pumped charge is seen to have a maximum as a function of barrier width. We find in Fig.~\ref{Fig:Tunn} (a) that this occurs when the width is just below the wavelength of the matter-waves. Quantum pumping using ``snow-plow'' dynamics can thus not occur efficiently if the width of the barrier differs significantly from the wavelength of the particles. For the case of $\Delta = 0$, the maximum occurs for narrower widths, and assuming analyticity of the scattering amplitudes, their derivatives should be continuous in the limit of vanishing or infinite width. Therefore based on Eq.~(\ref{Q_Line}) an almost identical argument as given above indicates that the pumped charge goes to zero in the limit $L\rightarrow 0,\infty$ and must therefore have a maximum at finite $L$.

\section{Summary and Conclusions}\label{Summary}
To our knowledge, this work is the first one suggesting an explicit implementation of ``snow-plow'' dynamics in an open geometry to produce quantum pumping. Earlier the mechanism had been suggested only for stirring \cite{CohenSnowPlow} in a closed circular geometry. In addition to this, we have shown that for finite pumping rates, Doppler shifted scattering gives a second contribution to the pumped charge. The two contributions (\ref{Q_Line}) and (\ref{Q_SP}) were derived using different assumptions, and are basically independent of each other, although both are based on the BTP-formula. It is here convenient to combine them since they are in some sense complementary in their dependence of the parameters, and in their different behavior.
If we consider the position $x_c(t)$ and its instantaneous velocity  $v_c(t)$ to be independent driving parameters, we can combine Eqs.~(\ref{Q_Line}) and (\ref{Q_SP}) into a single one for a line integral in the three-dimensional space spanned by ($x_c$,$v_c$,$V_0$) to give
\begin{equation}\label{ThreeDim}
Q = \frac{e}{\pi}\int_{\gamma}\vec{B}\cdot d\vec{X}_3,
\end{equation}
where $\gamma$ is the integration contour, $d\vec{X}_3 = (x_c,v_c,V_0)$, and the geometrical magnetic field vector $\vec{B}$ is given by
\begin{equation}\label{B_1}
B_1=-\frac{e}{\pi}k_F|r|^2,
\end{equation}
\begin{equation}\label{B_2}
B_2=t_-^*\frac{\partial t_-}{\partial v_c}+r_+^*\frac{\partial r_+}{\partial v_c},
\end{equation}
\begin{equation}\label{B_3}
B_3=t_-^*\frac{\partial t_-}{\partial V_0}+r_+^*\frac{\partial r_+}{\partial V_0}.
\end{equation}

\acknowledgments
The authors acknowledge the helpful advice of K.~K.~Das during early stages of this project.

\end{document}